\documentclass[aps,twocolumn,prd,nofootinbib,superscriptaddress]{revtex4}
\usepackage{amsmath}
\usepackage{graphicx}
\usepackage{dcolumn}
\usepackage{bm}
\usepackage{amssymb}
\usepackage{latexsym}

\newcommand{\be}{\begin{equation}}
\newcommand{\ee}{\end{equation}}
\newcommand{\bq}{\begin{eqnarray}}
\newcommand{\eq}{\end{eqnarray}}

\def\c{\mathfrak{c}}
\def\p{\mathfrak{p}}

\begin{document}


\title{Chaplygin inflation in loop quantum cosmology}

\author{Xin Zhang}
\affiliation{Department of Physics, College of Sciences,
Northeastern University, Shenyang 110004, China}
\author{Jingfei Zhang}
\affiliation{Department of Physics, College of Sciences,
Northeastern University, Shenyang 110004, China} \affiliation{School
of Physics and Optoelectronic Technology, Dalian University of
Technology, Dalian 116024, China}
\author{Jinglei Cui}
\affiliation{Department of Physics, College of Sciences,
Northeastern University, Shenyang 110004, China}
\author{Li Zhang}
\affiliation{Department of Physics, College of Sciences,
Northeastern University, Shenyang 110004, China}

\begin{abstract}
In this paper we discuss the inflationary universe in the context of
a Chaplygin gas equation of state within the framework of the
effective theory of loop quantum cosmology. Under the slow-roll
approximation, we calculate the primordial perturbations for this
model. We give the general expressions of the scalar spectral index,
its running, and the tensor-to-scalar ratio, etc. For the chaotic
inflation with a quadratic potential, using the WMAP 5-year results,
we determine the parameters of the Chaplygin inflation model in loop
quantum cosmology. The results are consistent with the WMAP
observations.
\end{abstract}

\maketitle

\section{Introduction}\label{sec:intro}

The inflation paradigm provides a compelling solution for many
long-standing problems of the cosmological standard model (such as
flatness, horizon, monopoles, etc.) by positing an epoch of
accelerated expansion in the early universe
\cite{Guth,inf2,inf3,inf4,Starobinsky:infl}. This accelerated period
of expansion also generates superhorizon fluctuations and thus
predicts an almost scale-invariant density perturbation power
spectrum, which has received strong observational support from the
measurement of the temperature fluctuation in the cosmic microwave
background (CMB) radiation
\cite{cobe,cobe2,wmap1,wmap1b,wmap1c,wmap3,wmap5}. Conceptually,
however, the inflationary scenario is incomplete due to the
existence of the big bang singularity \cite{Borde:2001nh}.
Einstein's classical theory of general relativity (GR) breaks down
near such a singularity since quantum effects are expected to be
important at very high energies in the early universe. Thus, the
classical theory of GR has to be replaced by some theoretical
framework of quantum gravity which should remain well defined even
at very high curvatures.

Loop quantum gravity (LQG) is a leading nonperturbative background
independent approach to quantizing gravity \cite{LQG,LQG2,LQG3}. The
underlying geometry in LQG is discrete and the continuum spacetime
can be obtained from the quantum geometry in a large eigenvalue
limit. Loop quantum cosmology (LQC) focuses on symmetry reduced
models (with homogeneous and isotropic space) but inherits
quantization scheme and techniques from LQG \cite{LQC}. Within the
framework of LQC, some long-standing issues concerning the quantum
nature of the big-bang are resolved in the context of homogeneous
and isotropic universe with a scalar field. Using extensive
analytical and numerical methods, the analysis of the evolution of
the semiclassical states for a spatially flat universe has shown
that the universe has a pre-big-bang branch, joined
deterministically to the post-big-bang branch by a quantum bounce in
the deep Planck regime through the LQC evolution
\cite{quantum1,quantum2,quantum3}. Thanks to the nonperturbative
background independent methods of LQC, the idea of the nonsingular
bounce can be realized in a natural fashion.

Recently, it has been shown that the discrete quantum dynamics can
be well approximated by an effective modified Friedmann dynamics
\cite{quantum2,quantum3,semiclass}. The modified Friedmann equation
can be obtained from the effective Hamiltonian constraint, which can
be used to investigate the role of nonperturbative quantum
correction conveniently. It is remarkable that the quantum geometric
effects lead to a $\rho^2$ modification to the Friedmann equation at
the scales when $\rho$ becomes comparable to a critical density
$\rho_c$ which is close to the Planck density ($\rho_c\approx 0.82
G^{-2}$) \cite{quantum2,quantum3,dualRS,geom,bounce}. Within the
framework of LQC, the inflationary universe model has been studied
in detail \cite{Zhang:2007bi}.

On the other hand, it is well known that the generalized Chaplygin
gas (GCG) is an alternative candidate for explaining the
acceleration of the universe. The GCG is described by an exotic
equation of state in the form \cite{gcg}
\begin{equation}
p_{ch}=-{A\over \rho_{ch}^\alpha},\label{gcg}
\end{equation}
where $\rho_{ch}$ and $p_{ch}$ are, respectively, the energy density
and pressure of the GCG, $\alpha$ is a constant in the range
$0<\alpha\leqslant 1$, and $A$ is a positive constant. It should
also be mentioned, however, that ``superluminal'' values $\alpha>1$
are well possible too, as was recently shown in Ref.
\cite{Gorini:2007ta}; causality requires only some change in Eq.
(\ref{gcg}) for $\rho_{ch}$ very close to $A^{1/\alpha}$. The case
$\alpha=1$ corresponds to the original Chaplygin gas \cite{cg,cg2}.
Inserting this equation of state into the stress-energy conservation
equation leads to the following energy density
\begin{equation}
\rho_{ch}=\left[A+{B\over a^{3(1+\alpha)}}\right]^{1\over
1+\alpha},\label{rho}
\end{equation}
where $a$ is the scale factor of the universe and $B$ is a positive
integration constant. The model of GCG as well as its further
generalization have been extensively studied in the literature
\cite{gcgexet,gcg2,gcg3,gcg4,gcg5,gcg6,gcg7,ngcg,mcg,mcg2,mcg3,mcg4}.
In addition, it should be pointed out that the GCG model may be
viewed as a modification of gravity \cite{Barreiro:2004bd}.

Recently, it was considered that the GCG model can also be used to
describe the early universe. In this consideration, Eq. (\ref{rho})
is not viewed as a consequence of the equation of state (\ref{gcg}),
but rather, as arising due to a modification of gravity itself.
Furthermore, one can also assume that during inflation the gravity
dynamics may give rise to a modified Friedmann equation
\cite{chapinflation}
\begin{equation}
3M_p^2H^2=\left[A+\rho_\phi^{(1+\alpha)}\right]^{1\over
1+\alpha},\label{chinf}
\end{equation}
where $M_p$ is the reduced Planck mass, $\rho_\phi$ is the energy
density of the inflaton field. This is the so-called Chaplygin
inspired inflation scenario \cite{chapinflation}, since the
modification in Eq. (\ref{chinf}) is realized from an extrapolation
of Eq. (\ref{rho}):
\begin{equation}
\rho_{ch}=\left[A+\rho_m^{(1+\alpha)}\right]^{1\over
1+\alpha}\longrightarrow
\left[A+\rho_\phi^{(1+\alpha)}\right]^{1\over 1+\alpha},
\end{equation}
where $\rho_m\sim a^{-3}$ corresponds to the matter energy density.
Following this idea, lately, some work has been done, this involves
tachyon-Chaplygin inflation \cite{tachyon-chaplygin}, warm-Chaplygin
inflation \cite{warmchap}, Chaplygin inflation on a brane
\cite{chapbrane}, and so forth \cite{Monerat:2007ud}. In this paper,
we shall study the Chaplygin inspired inflationary model in loop
quantum cosmology.

This paper is organized as follows. In the next section we briefly
review the effective theory of LQC. In section \ref{sec:lqc}, we
describe the Chaplygin inflationary universe in the framework of
LQC. In section \ref{sec:perturb}, we study the primordial
perturbations and give the expressions of scalar and tensor
perturbations for this model. Also, we shall compare our results to
the WMAP five-year observations. Conclusion is given in section
\ref{sec:concl}.

\section{Effective dynamics in loop quantum
cosmology}\label{sec:lqc}

LQG is a canonical quantization of gravity based upon
Ashtekar-Barbero connection variables. The phase space of classical
GR in LQG is spanned by SU(2) connection $A_a^i$ and the triad
$E_i^a$ on a 3-manifold $\cal{M}$ (labels $a$ and $i$ denote space
and internal indices respectively), which are two conjugate
variables encoding curvature and spatial geometry, respectively.
Likewise, LQC is a canonical quantization of homogeneous spacetimes
based upon techniques used in LQG. In LQC, due to the symmetries of
the homogeneous and isotropic spacetime, the phase space structure
is simplified, i.e., the connection is determined by a single
quantity labeled $\c$ and likewise the triad is determined by a
parameter $\p$. The variables $\c$ and $\p$ are canonically
conjugate with Poisson bracket $\{\c, \p\}=\gamma/3$, here we have
used the unit $M_p=1$ for convenience, and $\gamma$ is the
dimensionless Barbero-Immirzi parameter which is set to be
$\gamma\approx 0.2375$ by the black hole thermodynamics in LQG. For
the spatially flat model of cosmology, the new variables have the
relations with the metric components of the
Friedmann-Robertson-Walker (FRW) universe as
\begin{equation}
\c=\gamma\dot{a},~~~~\p=a^2,\label{newvars}
\end{equation}
where $a$ is the scale factor of the universe. Classically in terms
of the connection-triads variables the Hamiltonian constraint is
given by
\begin{equation}
{\cal H}_{\rm cl}=-{3\sqrt{\p}\over \gamma^2}\c^2+{\cal H}_{\rm M},
\end{equation}
where ${\cal H}_{\rm M}$ is the matter Hamiltonian.

The elementary variables used for quantization in LQC are the triads
and holonomies of the connection. The holonomy over an edge of a
loop is defined as $h_i(\mu)=\cos (\mu\c/2)+2\sin (\mu\c/2)\tau_i$,
where $\tau_i$ is related to Pauli spin matrices as
$\tau_i=-i\sigma_i/2$ and dimensionless $\mu$ is related to the
physical length of the edge over which holonomy is evaluated (note
that $\mu$ is also the eigenvalue of the triad operator $\hat{\p}$).
In the Hamiltonian formulation for homogeneous and isotropic
spacetime, the dynamical equations can be determined completely by
the Hamiltonian constraint. Under quantization, the Hamiltonian
constraint gets promoted to an operator and the quantum wave
functions are annihilated by the operator of the Hamiltonian
constraint. In LQC, it is expected that modifications due to LQC
effects will appear in the Hamiltonian constraint, and from the
modified Hamiltonian constraint the effective Friedmann constraint
will be derived. In quantization the Hamiltonian constraint operator
is obtained by promoting the holonomies and the triads to the
corresponding operators. Consequently, this leads to a discrete
quantum difference equation, which indicates that the underlying
geometry in LQC is discrete \cite{solsingu1}. Interestingly, the
solutions of this difference equation are nonsingular.

So far we see that the underlying dynamics in LQC is governed by a
discrete quantum difference equation in quantum geometry. However,
an effective Hamiltonian description on a continuum spacetime can be
constructed by using semiclassical states, which has been shown to
very well approximate the quantum dynamics \cite{quantum2,quantum3}.
This analysis reveals that on backward evolution of our expanding
phase of the universe, the universe bounces at a critical density
(near the big bang singularity) into a contracting branch
\cite{quantum1,bounce}. Thus the classical singular problem can be
successfully overcome within the context of LQC by a nonsingular
bounce. In addition, the effective equations for the modified
Friedmann dynamics can be derived from the effective Hamiltonian
constraint with loop quantum modifications, which can be used to
investigate the role of nonperturbative quantum corrections. An
important feature for the modified dynamics is that a $\rho^2$ term
which is relevant in the high energy regime is included in the
classical Friedmann equation. The modified term is negative definite
implying a bounce when the energy density reaches a critical value
on the order of the Planck density.

The effective Hamiltonian constraint, to leading order, is given by
\cite{semiclass}
\begin{equation}
{\cal H}_{\rm eff}=-{3\over \gamma^2\bar{\mu}^2}a
\sin^2(\bar{\mu}\c)+{\cal H}_{\rm M},
\end{equation}
where $\bar{\mu}$ is the kinematical length of the edge of a square
loop which has the area given by the minimum eigenvalue of the area
operator in LQG; the area is ${\cal A}=\bar{\mu}^2 a^2=\alpha
l_{p}^2$, where $\alpha$ is of the order unity and $l_{p}$ is the
Planck length. The modified Friedmann equation can then be derived
by using the Hamilton's equation for $\p$,
\begin{equation}
\dot{\p}=\{\p, {\cal H}_{\rm eff}\}=-{\gamma\over 3}{\partial {\cal
H}_{\rm eff}\over \partial
\c}={2a\over\gamma\bar{\mu}}\sin(\bar{\mu}\c)\cos(\bar{\mu}\c),
\end{equation}
which combined with Eq. (\ref{newvars}) yields the rate of change of
the scale factor
\begin{equation}
\dot{a}={1\over
\gamma\bar{\mu}}\sin(\bar{\mu}\c)\cos(\bar{\mu}\c).\label{dota}
\end{equation}
Furthermore, the vanishing of the Hamiltonian constraint, ${\cal
H}_{\rm eff}\approx 0$, implies
\begin{equation}
\sin^2(\bar{\mu}\c)={\gamma^2\bar{\mu}^2\over 3a}{\cal H}_{\rm
M}.\label{sin}
\end{equation}
Combining Eqs. (\ref{dota}) and (\ref{sin}) yields the effective
Friedmann equation for the Hubble rate $H=\dot{a}/a$,
\begin{equation}
H^2={1\over3}\rho\left(1-{\rho\over\rho_c}\right),\label{Feq}
\end{equation}
with the critical density given by
\begin{equation}
\rho_c=4\sqrt{3}\gamma^{-3}.\label{crit}
\end{equation}
The modified Friedmann equation provides an effective description
for LQC which very well approximates the underlying discrete quantum
dynamics. The nonperturbative quantum geometric effects are
manifested in the modified Friedmann equation with a $\rho^2$
correction term. The negative definition of the $\rho^2$ term
implies that the Hubble parameter vanishes when $\rho=\rho_c$ and
the universe experiences a turnaround in the scale factor. When
$\rho\ll \rho_c$, the modifications to the Friedmann equation become
negligible, and the standard Friedmann equation is recovered. In
addition, it should be noted that, interestingly, $\rho^2$
modifications also appear in string inspired braneworld scenarios
and it has been shown that there exist interesting dualities between
the two frameworks \cite{dualRS}. Such modifications in braneworlds,
however, are usually positive definite so that a bounce is absent,
unless the existence of two timelike extra dimensions is assumed
\cite{bouncingbrane}. For extensive studies in this framework, see
e.g. \cite{lqcexet,lqce2,lqce3,lqce4,lqce5,lqce6,lqce7,lqce8}. Other
mechanism of realizing a bouncing (or oscillating) universe can be
found in e.g. \cite{cycl,cycl2,cycl3,cycl4,cycl5,cycl6}.

\section{Chaplygin inflationary universe in loop quantum
cosmology}\label{sec:chapinfl}

In this section, we shall investigate Chaplygin inspired inflation
within the framework of the effective theory of LQC. In this case
the dynamics of the inflationary universe is governed by the
following equation,
\begin{equation}
3H^2=\left(A+\rho_\phi ^{(1+\alpha)}\right)^{1\over
1+\alpha}\left[1-{(A+\rho_\phi ^{(1+\alpha)})^{1\over 1+\alpha}\over
\rho_c}\right].\label{chapfeq}
\end{equation}
Here, $\rho_\phi$ is the canonical inflaton,
$\rho_\phi=\dot{\phi}/2+V(\phi)$, and $V(\phi)$ is its potential. It
should be noted that in the low-energy regime, $[A+\rho_\phi
^{(1+\alpha)}]^{1/(1+\alpha)}\ll \rho_c$, the standard Chaplygin
inflation model is recovered; while in the high-energy regime, the
quantum gravity effects leads to a super-inflation that is in the
``primary inflation''. However, what is of interest for us is the
``observable inflation'' with the last 60 $e$-foldings. We thus only
discuss the subsequent normal inflation stage (after the
super-inflation phase) in what follows. Also, in the following we
will take $\alpha=1$ for simplicity, which means the usual Chaplygin
gas.

For the inflaton field, its equation of motion is of the standard
form,
\begin{equation}
\ddot{\phi}+3H\dot{\phi}+V'=0,\label{eom}
\end{equation}
where dots denote the derivatives with respect to the cosmic time
and $V'=dV(\phi)/d\phi$, and the Hubble parameter $H$ is given by
Eq. (\ref{chapfeq}). If $\dot{\phi}^2\ll V(\phi)$ and
$\ddot{\phi}\ll 3H\dot{\phi}$, the scalar field will slowly roll
down its potential, and the exact evolution equation (\ref{eom}) can
be replaced by the slow-roll approximation
\begin{equation}
\dot{\phi}=-V'/3H.\label{slow}
\end{equation}
Under the slow-roll approximation, the energy density of the scalar
field approximates as $\rho\sim V(\phi)$, the Friedmann equation
(\ref{chapfeq}) consequently can be written as
\begin{equation}
H^2={1\over 3} \sqrt{A+V^2}\left(1-{\sqrt{A+V^2}\over
\rho_c}\right).\label{Feq2}
\end{equation}
We can also define the slow-roll parameters
\begin{equation}
\epsilon=-{\dot{H}\over H^2}= {1\over 2}{VV'2\over
(A+V^2)^{3/2}}{(1-{2\sqrt{A+V^2}\over \rho_c})\over
(1-{\sqrt{A+V^2}\over \rho_c})^2},\label{epsilon}
\end{equation}
\begin{equation}
\eta=\epsilon+\delta= {V''\over
(A+V^2)^{1/2}}\left(1-{\sqrt{A+V^2}\over
\rho_c}\right)^{-1},\label{eta}
\end{equation}
where $\delta=-\ddot{\phi}/H\dot{\phi}$, by definition. The
slow-roll condition can be expressed as $\epsilon, ~|\eta|\ll 1$.

Note that in the limit $A\rightarrow 0$, the slow-roll parameters
$\epsilon$ and $\eta$ coincide with the case in inflation model in
LQC \cite{Zhang:2007bi}. Also, in the low-energy limit,
$\sqrt{A+\rho_\phi ^2}\ll \rho_c$, the slow-roll parameters reduce
to the standard form of Chaplygin inflation.

The parameter $\epsilon$ quantifies how much the Hubble rate $H$
changes with time during inflation. Inflation can be attained only
if $\epsilon<1$. This condition could be written explicitly in terms
of the scalar potential $V$ and its derivative $V'$
\begin{equation}
\begin{array}{l}
VV'2\left[1-{2(A+V^2)^{1/2}\over\rho_c}\right]\\
< (A+V^2)^{3/2}\left[1-{(A+V^2)^{1/2}\over\rho_c}\right]^2.
\end{array}
\end{equation}
Inflation ends when $\epsilon<1$ ceases to be satisfied, which
implies
\begin{equation}
\begin{array}{l}
V_fV_f'2\left[1-{2(A+V_f^2)^{1/2}\over\rho_c}\right]\\ =
(A+V_f^2)^{3/2}\left[1-{(A+V_f^2)^{1/2}\over\rho_c}\right]^2,
\end{array}
\end{equation}
where the subscript $f$ is used to mark the end of inflation.

The small change of $e$-foldings satisfies $dN\equiv -Hdt$; using
Eqs. (\ref{slow}) and (\ref{Feq2}), one obtains the number of
$e$-foldings of slow-roll inflation remaining at a given epoch,
\begin{equation}
N(\phi_*)=\int\limits_{\phi_f}^{\phi_*}{\sqrt{A+V^2}\over
V'}{\left[1-{\sqrt{A+V^2}\over \rho_c}\right]}d\phi,\label{efold}
\end{equation}
where the subscript $*$ marks the epoch when the cosmological scales
exit the horizon.

\section{Primordial perturbations and WMAP five-year
results}\label{sec:perturb}

In this section we shall study the primordial perturbations of this
inflationary model. We shall give the expressions of scalar and
tensor perturbations for the Chaplygin inflation in LQC.

The perturbation $\delta\phi$ can be treated as a massless free
field, and its vacuum fluctuation can be regarded as a classical
quantity a few Hubble times after horizon exit. The spectrum of the
field perturbation is
\begin{equation}
\Delta^2_{\delta\phi}=(H/2\pi)^2.\label{Pphi}
\end{equation}
The corresponding curvature perturbation is given by ${\cal
R}=(H/\dot{\phi})\delta\phi$. Using Eqs. (\ref{slow}), (\ref{Feq2})
and (\ref{Pphi}), we get the amplitude of scalar perturbation as
\begin{equation}
\Delta^2_{\cal R}=\left({H^2\over
2\pi{\dot{\phi}}}\right)^2={(A+V^2)^{3/2}\over 12\pi^2
V'^2}\left(1-{\sqrt{A+V^2}\over \rho_c}\right)^3,\label{PR}
\end{equation}
which is evaluated at the Hubble radius crossing $k=aH$. Note that
in the limit $A\rightarrow 0$, the amplitude of the scalar
perturbation given by Eq. (\ref{PR}) coincides with Ref.
\cite{Zhang:2007bi}.

Since $H$ is slowly varying, we have the relation $d\ln k(\phi)= -d
N(\phi)$. Then, from Eq. (\ref{PR}) we get the scalar spectral
index, $n_s-1=d\ln \Delta^2_{\cal R}/d\ln k=-6\epsilon+2\eta$, or
equivalently,
\begin{eqnarray}
n_s&=&1-(A+V^2)^{-1/2}\left[1-{(A+V^2)^{1/2}\over
\rho_c}\right]^{-1}\nonumber\\
 &\times &\left({3VV'^2\over
(A+V^2)}{[1-{2(A+V^2)^{1/2}\over\rho_c}]\over
[1-{(A+V^2)^{1/2}\over\rho_c}]}\right).\label{ns}
\end{eqnarray}
Once again, we notice that in the limit $A\rightarrow 0$ the scalar
spectral index coincides with that in Ref. \cite{Zhang:2007bi}.

Another important quantity for inflationary model is the running of
the scalar spectral index $\alpha_s=dn_s/d\ln k$. From Eq.
(\ref{ns}), we can get the running of the scalar spectral index for
this model
\begin{equation}
\alpha_s=\left({4(A+V^2)\over
VV'}\right)\left[{1-{(A+V^2)^{1/2}\over\rho_c}\over
1-{2(A+V^2)^{1/2}\over\rho_c}}\right][3\epsilon_{,\phi}-\eta_{,\phi}]\epsilon,\label{as}
\end{equation}
where $\epsilon$ and $\eta$ are given by Eqs. (\ref{epsilon}) and
(\ref{eta}), and the subscript $_{,\phi}$ labels the derivative with
respect to the inflaton field $\phi$.

Inflation also generates gravitational waves with two independent
components $h_{+,\times}$ which have the same action as a massless
scalar field. Likewise, the amplitude of tensor perturbation can
also be given,
\begin{equation}
\Delta^2_h=8\left({H\over 2\pi}\right)^2={2(A+V^2)^{1/2}\over
3\pi^2}\left(1-{\sqrt{A+V^2}\over \rho_c}\right),\label{PT}
\end{equation}
which is also evaluated at the horizon exit $k=aH$. The spectral
index of tensor perturbation is given by $n_t=d\Delta^2_h/d\ln k$
that will not be written explicitly here. Then, from the expressions
(\ref{PR}) and (\ref{PT}), we can write the tensor-to-scalar ratio,
at $k=aH$, as
\begin{equation}
r={\Delta^2_h\over\Delta^2_{\cal R}}={8V'^2\over
(A+V^2)[1-(A+V^2)^{1/2}/\rho_c]}.
\end{equation}

Recently, the Wilkinson Microwave Anisotropy Probe (WMAP) five-year
data were released \cite{wmap5}. The WMAP 5-year data provide
stringent constraints on inflationary models. If we assume that the
primordial fluctuations are adiabatic with a power law spectrum, the
WMAP 5-year data (WMAP only) give: $n_s=0.963^{+0.014}_{-0.015}$;
while the WMAP data combined with the distance measurements from the
Type Ia supernovae (SN) and Baryon Acoustic Oscillations (BAO) in
the distribution of galaxies (WMAP+BAO+SN) give:
$n_s=0.960^{+0.014}_{-0.013}$. The amplitude of curvature
perturbation is $\Delta^2_{\cal R}=2.4\times 10^{-9}$, measured by
WMAP at $k_*=0.002~{\rm Mpc}^{-1}$. With the WMAP data combined with
BAO and SN, it is found that the limit on the tensor-to-scalar ratio
is $r<0.20$ ($95\%$ CL) and $n_s>1$ is disfavored even when
gravitational waves are included. We will make use of the WMAP
5-year results to determine the parameters of the model.

Consider now an inflaton scalar field $\phi$ with a chaotic
potential. We focus on a quadratic potential $V(\phi)=m^2\phi^2/2$,
where $m$ is the mass of the scalar field. First, we should evaluate
$\phi_*$ (that corresponds to the time of horizon-crossing) by using
Eq. (\ref{efold}). In view of that the LQC effect is not prominent
(for convenience we define $\nu\equiv (A+V^2)/\rho_c$, and we have
$\nu<10^{-9}$, see \cite{Zhang:2007bi}), we neglect $\nu$ in this
evaluation. Then, the integral in Eq. (\ref{efold}) can be given
directly (see Eq. (19) in \cite{chapinflation}). Furthermore,
considering $\phi_f\sim 0$ and the unimportance of logarithm term,
one obtains $\phi_*=2\sqrt{N}$ \cite{chapinflation}.

Next, using the WMAP 5-year results (WMAP+BAO+SN), $\Delta^2_{\cal
R}=2.4\times 10^{-9}$ and $n_s=0.960$, we can determine the
parameters of the LQC Chaplygin inflation model. Taking $N=55$, we
derive $A=3.36\times 10^{-18}$ and $m=5.9\times 10^{-6}$, note that
here the unit $M_p=1$ has been used. Based on these values of
parameters, we can further evaluate the other observational
quantities. For example, we give the running of the scalar spectral
index $\alpha_s=-6.8\times 10^{-5}$ and the tensor-to-scalar ratio
$r=0.1$, which are in good consistence with the WMAP observations.
Also, we verified the fact that the LQC effect can only leads to a
very tiny imprint in the primordial power spectrum, $\nu=8.5\times
10^{-10}$, which is in good accordance with the result in Ref.
\cite{Zhang:2007bi}.

\section{conclusion}\label{sec:concl}

In this paper we have studied the Chaplygin inflation model in the
framework of the effective theory of LQC. In LQC, the
nonperturbative quantum geometry effects lead to a $\rho^2$ term
with a negative sign in the modified Friedmann equation. On the
other hand, inspired by the Chaplygin equation of state, a
phenomenological modification of gravity is also considered in the
Friedmann dynamics. The LQC Chaplygin inflationary model combines
the above both features.

Under the slow-roll approximation, we calculated the primordial
perturbations for this model. We gave the general expressions of the
scalar spectral index $n_s$, its running $\alpha_s$, and the
tensor-to-scalar ratio $r$, etc. For the chaotic inflation with a
quadratic potential, using the WMAP 5-year results, $\Delta^2_{\cal
R}(k_*)=2.4\times 10^{-9}$ and $n_s(k_*)=0.960$, we determined the
parameters of the LQC Chaplygin inflation model, $A=3.36\times
10^{-18}$ and $m=5.9\times 10^{-6}$. This leads to the results of
the running of the scalar spectral index $\alpha_s(k_*)=-6.8\times
10^{-5}$ and the tensor-to-scalar ratio $r(k_*)=0.1$, which are
consistent with the WMAP observations.

The theory of LQC gives rise to a quantum bounce in the high energy
regime when the loop quantum effects are dominative power. The
classical big-bang singularity is thus replaced by the quantum
bounce. After the bounce, a super-inflation phase was emergent in a
natural way. Then the universe underwent a normal inflation stage.
In this paper, we restrict that the Chaplygin-inflation happens in
the normal inflation phase. This leads to that the imprint of the
LQC effect in CMB sky is too weak to be observed by present
observational data (the LQC parameter $\nu<10^{-9}$, see Ref.
\cite{Zhang:2007bi}). For the Chaplygin inflation case with
quadratic potential, we derived $\nu=8.5\times 10^{-10}$ that is in
agreement with the conclusion of Ref. \cite{Zhang:2007bi}. If we set
the Chaplygin inflation to happen in the super-inflation phase, it
is believed that the LQC effect should be prominent and the imprint
of LQC effect may be detected in the CMB sky. We hope to return to
this point in the near future.

\acknowledgments

This work was supported in part by the Natural Science Foundation of
China.


\end{document}